\documentclass[12pt]{article}
\usepackage[english]{babel}
\usepackage{epsfig}
\begin{document}
\large

\par
\noindent {\bf Important remarks to Wolfenstein's equation for
passing neutrino through the matter}

\par
\begin{center}
\vspace{0.3cm} Kh. M. Beshtoev
\par
\vspace{0.3cm} Joint Institute for Nuclear Research, Joliot Curie
6, 141980 Dubna, Moscow region, Russia.
\end{center}
\vspace{0.3cm}

\par
Abstract\\

\par
It is supposed that while neutrino passing through the matter a
resonance enhancement of neutrino oscillations in the matter
appears. It is shown that Wolfenstein's equation, for neutrino
passing through the matter, contains a disadvantage (does not take
into account the law of momentum conservation). It leads, for
example, to changing of the effective mass of neutrino by the
value of $0.87 \cdot 10^{-2} eV$ from the very small value of the
energy polarization of the matter caused by neutrino which is
equal to $5 \cdot 10^{-12} eV$. After removing this disadvantage
(i.e., taking into account this law) we have obtained a solution
of this equation. In this solution a very small enhancement of
neutrino oscillations in the matter appears due to the smallness
of the energy polarization of the matter caused by neutrino.

\section{Introduction}

\par
The suggestion that, in analogy with $K^{o}, \bar K^{o}$
oscillations, there could be neutrino-antineutrino oscillations (
$\nu \rightarrow \bar \nu$), was made by Pontecorvo \cite{1} in
1957. It was subsequently considered by Maki et al. \cite{2} and
Pontecorvo \cite{3} that there could be mixings (and oscillations)
of neutrinos of different flavors (i.e., $\nu_{e} \rightarrow
\nu_{\mu}$ transitions).
\par
The first experiment \cite{4} on the solar neutrinos has shown
that there is a deficit of neutrinos, i.e., the solar neutrinos
flux detected in the experiment was few times smaller than the
flux computed in the framework of the Sun Standard Model \cite{5}.
The subsequent experiments and theoretical computation have
confirmed the deficit of the solar neutrinos \cite{6}.
\par
The short base reactor and accelerator experiments \cite{7} have
shown that there is no neutrino deficit, then this result was
interpreted as an indication that the neutrino vacuum angle mixing
is very small (subsequent experiments have shown \cite{8} that
this vacuum angle is big and near the maximal value). Then the
question appears: what is the deficit of the solar neutrinos
related with? In 1978 the work by L. Wolfenstein \cite{9} appeared
where an equation describing neutrino passing through the matter
was formulated (afterwards that equation was named Wolfenstein's).
In the framework of this equation the enhancement of neutrino
oscillations in the matter arises via weak interactions (critical
remarks to this equation see in \cite{10}). This mechanism of
neutrino oscillations enhancement in the matter attracted the
attention of neutrino physicists after publications \cite{11} by
S. Mikheyev and A. Smirnov where it was shown that in the
framework of this equation the resonance enhancement of neutrino
oscillations in the matter would take place. Also it is clear that
the adiabatic neutrino transitions can arise in the matter if
effective masses of neutrinos change in the matter \cite{12}.
\par
This work is devoted to discussion of neutrino oscillations in the
matter by using the Wolfenstein's type equation.

\section{The neutrino (particle) passing through the matter}

\par
Before consideration of a neutrino (particle) passing through the
matter it is necessary to gain some understanding of the physical
origin of this mechanism. While neutrino passing through the
matter there can be two processes- neutrino scattering and
polarization of the matter by neutrino. Our interest is related
with neutrino elastic interactions in the matter namely with
neutrino forward elastic scattering, i.e., polarization of the
matter by the passing neutrino. The neutrino passing through the
matter at its forward scattering can be considered by using the
following Wolfenstein's equation \cite{9}
$$
i \frac{d\nu_{Ph}}{dt} = (\hat E + \hat W) \nu_{Ph} \equiv (
\sqrt{p^2 I + {\hat M}^2} + \hat W ) \nu_{Ph} , \eqno(1)
$$
where $p, \hat M^{2}, \hat W_i $ are, respectively, the momentum,
the (nondiagonal) square mass matrix in vacuum, and  the matrix,
taking  into account neutrino interactions in the matter,
$$
\nu_{Ph} = \left (\begin{array}{c} \nu_{e}\\
\nu_{\mu} \end{array} \right) , \qquad \hat I = \left(
\begin{array}{cc} 1&0\\0&1 \end{array} \right) ,
$$
$$
\hat M^{2} = \left( \begin{array}{cc} m^{2}_{\nu_{e}\nu_{e}}&
m^{2}_{\nu_{e} \nu_{\mu}}\\ m^{2}_{\nu_{\mu}\nu_{e}}&
m^{2}_{\nu_{\mu} \nu_{\mu}} \end{array} \right) \qquad \hat W =
\left(
\begin{array}{cc} \hat W_e & 0 \\ 0 & \hat W_\mu \end{array}
\right). \eqno(2)
$$
This equation has standard solution which was found by S. P.
Mikheyev and A. Ju. Smirnov \cite{11} which come to resonance
enhancement of neutrino oscillations in the matter. Now come to
consider this decision.

\subsection{The standard mechanism of resonance enhancement
of neutrino oscillations in the matter and some critical remarks}

\par
In the ultrarelativistic limit ($E \simeq p\hat I + \frac{ {\hat
M}^2}{2p}$), the evolution equation for the neutrino wave function
$\nu_{Ph}$ in the matter has the following form \cite{9},
\cite{11}:
$$
i \frac{d\nu_{Ph}}{dt} = ( p\hat I + \frac{ {\hat M}^2}{2p} + \hat
W ) \nu_{Ph} , \eqno(3)
$$
where $p, \hat M^{2}, \hat W_i $ are, respectively, the momentum,
the (nondiagonal) square mass matrix in vacuum, and  the matrix,
taking  into account neutrino interactions in the matter.
\par
If we suppose that neutrinos in the matter behave analogously to
the photon in the matter (i.e., the polarization appears while
neutrino passing through the matter) and the neutrino refraction
indices are defined by the following expression:
$$
n_{i} = 1 + \frac{2 \pi N}{p^{2}} f_{i}(0) = 1 + 2 \frac{\pi
W_i}{p} , \eqno(4)
$$
where $i$ is a type of neutrinos $(e, \mu, \tau)$, $N$ is density
of the matter, $f_{i}(0)$ is a real part of the forward scattering
amplitude, then $W_i$ characterizes the polarization of the matter
by neutrinos (i.e. it is the energy of the matter polarization).
In reality, as we will see below, there is fundamental difference:
photon is a massless particle while neutrino is a massive particle
and this distinction is fundamental.
\par
The electron neutrino ($\nu_{e}$)  in the matter interacts via
$W^{\pm}, Z^{0}$ bosons and $\nu_{\mu}, \nu_{\tau}$ interact only
via $Z^{0}$ boson. These differences in interactions lead to the
following differences in the refraction coefficients of $\nu_{e}$
and $\nu_{\mu}, \nu_{\tau}$:

$$
\Delta n = \frac{2 \pi {\it n_e}}{p^{2}} \Delta f(0) , \eqno(5)
$$
$$
\Delta f(0) =  \sqrt{2} \frac{G_F}{2 \pi} p ,
$$
$$
E_{\rm{eff}} = \sqrt{p^2 + m^2} + \langle{e} \nu|H_{\rm{eff}}|e
\nu\rangle \approx p +\frac{m^2}{2 p} + \sqrt{2} G_F {\it n_e},
$$
where $G_F$ is the Fermi constant.
\par
Energy of the matter polarization $E$ is
$$
E \approx W = \sqrt{2} G_F {\it n_e} ,\quad W  = 7.6  \left(
\frac{n_e}{n_0} \right) \cdot 10^{-14} eV. \eqno(5')
$$
where $G_F$ is the Fermi constant, $n_e$ is the electron density
in the matter. For the Sun
$$
E^{\rm{SUN}} \approx 10^{-13}\div 10^{-11}\,{\rm{eV}}. \eqno(5'')
$$
\par
Therefore the velocities (or effective masses) of $\nu_{e}$ and
$\nu_{\mu}, \nu_{\tau}$ in the matter are different. And at the
suitable density of the matter this difference can result in
resonance enhancement of neutrino oscillations in the matter
\cite{9}, \cite{12}. Expression for $\sin ^{2} 2\theta _{m}$ in
the matter has the following form:
\par
$$
\sin ^{2} 2\theta _{m} = \sin^{2} 2\theta \cdot [(\cos 2\theta  -
{L_{0}\over L^{0}})^{2} + \sin ^{2} 2\theta ]^{-1} , \eqno(6)
$$
where $\sin ^{2} 2\theta _{m}$ and $\sin^{2} 2\theta$ characterize
neutrino mixings in the matter and vacuum, $L_{0}$ and $L^{0}$ are
lengths of neutrino oscillations in vacuum and neutrino refraction
length in the matter:
$$
L_{0} = \frac{4 \pi E_{\nu} \hbar}{\Delta m^2 c^3} \qquad L^{0} =
\frac{\sqrt{2} \pi \hbar c}{G_F n_e}, \eqno(7)
$$
where $E_{\nu}$ is neutrino energy, $\Delta m^2 = m_2^2 - m_1^2$ -
difference between squared neutrino masses, $c$ is light velocity,
$\hbar$ is Plank constant, $G_F$ is Fermi constant and $n_e$ is
electron density of the matter.
\par
Probability of $\nu_e \to \nu_\mu$ neutrino transitions is given
by the following expression ($E \simeq p c$):
$$
P(E, t, ...) = 1 - \sin ^{2} 2\theta _{m} \sin  {2\pi c t\over
L_{m}} , \eqno(8)
$$
where $ L_{m} = {\sin  2\theta _{m}\over \sin  2\theta } L_o$.
\par
At resonance
$$ \cos  2\theta  \cong  {L_{0}\over
L^{0}}\qquad sin^{2} 2\theta_{m} \cong 1\qquad \theta_{m} \cong
{\pi \over 4} . \eqno(9)
$$
The expression (9) for resonance condition can be rewritten in the
following form:
$$
\sqrt{2} G_F n_e = \frac{\Delta m^2}{2 E^{res}_{\nu_e}} cos 2
\theta , \eqno(10)
$$
or
$$
E^{res}_\nu = \frac{\Delta m^2 cos 2 \theta}{2 W } \quad \to
\Delta m^2 - \frac{2 E^{res}_\nu W}{cos 2 \theta} = 0 . \eqno(11)
$$
If we consider $\nu_e \to \nu_\mu$ and use KamLAND data
\cite{13kamland}
$$
tan^2 \theta_{1 2} = 0.56 (+0.10, - 0.07) (stat) (+0.1, - 0.06)
(syst), \quad \theta = 36.8^o,
$$
$$
\Delta m^2_{1 2} = 7.58 (+0.14, -0.13)(stat)\pm 0.15 (syst)\times
10^{-5} eV^2, \eqno(12)
$$
then at $n_e = 65.8 n_o$ energy $W^{Sun}$ of neutrino polarization
is $W^{Sun} = 5\times 10^{-12} eV$ and for $ E^{res}_\nu$ we
obtain
$$
E^{res}_\nu = 2.14 \times 10^{6} eV = 2.14 MeV . \eqno(13)
$$
The expressions (9)-(13) mean that when the electron neutrino with
energy $E_\nu = 2.14 \quad MeV$ is passing through the Sun matter,
the effective mass of electron neutrino becomes equal to the muon
neutrino mass (see below expressions (23),(23')) and as result
there is resonance transition of electron neutrinos into muon
neutrinos. In this case changing of the squared mass of neutrino
$\nu_1$ is $\Delta m^2_{1 2} = 7.58 \times 10^{-5} eV^2$ (we
suppose that $m_2 > m_1$), i. e., effective mass of neutrino
$\nu_1$ is
$$
m_{1, eff}^2 \simeq m_1^2 + 7.58 \times 10^{-5} eV^2, \eqno(14)
$$
and (in reality $m^{matt}_{\nu_e} \simeq m_{\nu_\mu}$, see
expression (23),(24))
$$
m_{1, matt}^2 \approx m_2^2 . \eqno(15)
$$
We see that this additional big mass arises at polarization of the
matter by neutrino with energy $W = 5\times 10^{-12} eV$! It is a
very strange result. A primary ultrarelativistic electron neutrino
having energy $E_\nu = 2.14 \times 10^{6} eV$ interacts with the
matter with the energy $W = 5 \times 10^{-12} eV$ and as a result
we obtained the mass increase on $\delta m \approx \sqrt{7.58
\times 10^{-5}} = 0.87 \times 10^{-2} eV$. We know that the matter
polarization has to move with the velocity equal to the neutrino
velocity which generates this polarization. Then the energy of the
electron neutrino has to increase on
$$
\Delta E_\nu \approx \delta m \gamma, \eqno(16)
$$
where $\gamma = \frac{E_\nu}{m_\nu}$ and neutrino velocity $v
\simeq c$.  Why did we come to this result? It is a consequence of
the above used approach when we included the full energy of the
matter polarization in the neutrino mass. It is possible only at a
serious violation of the law of energy-momentum conservation. We
can suppose that increasing of neutrino effective mass is
accompanied by the decreasing of neutrino velocity, i.e., the
energy is conserved but it is not save the situation since the
mass increasing is $\delta m \approx \sqrt{7.58 \times 10^{-5}} =
0.87 \times 10^{-2} eV$ while the energy of matter polarization is
$W = 5 \times 10^{-12} eV$). It means that the mechanism of
resonance enhancement of neutrino oscillations in the matter can
be realized at violation the law of energy-momentum conservation.
The approach which was considered above becomes physically
realizable only when $p^2 << M^2$ and then full energy of the
matter polarization is transformed in neutrino mass, but the
neutrinos produced in weak interactions are relativistic since the
neutrino masses are very small.
\par
As it was stressed above the approach which was used should work
in a non-relativistic case when $p^2 << M^2$ but not in the
ultrarelativistic case. Then expression for neutrino energy will
have the form
$$
\sqrt{p^2 + M^2} + W = \sqrt{p^2 + M'^2} \to M'^2 = M^2 + 2 W
\sqrt{p^2 + M^2} + W^2, \eqno(17)
$$
then taking into account that $p^2 << M^2$, $W^2 << M^2$ we obtain
$$
M'^2 \simeq (M + W)^2. \eqno(18)
$$
Then the Wolfenstein's equation (1) can be written in the
following form:
$$
i \frac{d\nu_{Ph}}{dt} = \sqrt{p^2 I + {(\hat M + \hat W)}^2}
\nu_{Ph} \equiv \sqrt{p^2 I + (\hat M')^2} \nu_{Ph} \simeq (\hat
M' + \frac{{\hat p}^2}{2 M'^2})\nu_{Ph}, \eqno(19)
$$
Then we see that, at low energies we can include the full energy
$W$ of the matter polarization in the neutrino mass and
expressions $(6) \div (11)$ will be replaced by expressions
$(17)\div(19)$. Then by diagonalization of mass the matrix $M'$ we
obtain that the neutrino mixing angle change and at appropriate
conditions (at enough large value of $W$) there will take place
resonance enhancement of neutrino oscillations in the matter. It
is necessary to remark that the full energy of matter polarization
$E^{Sun} \approx 10^{-13}\div 10^{-11}\,{\rm{eV}}$ which arise at
passing electron neutrino through the Sun is too small to generate
a resonance enhancement of neutrino oscillations. Above we
supposed that masses can be generate in the framework of the
standard weak interactions, i.e., there is no problem with mass
generation.
\par
Now we consider common case where the total energy of the matter
polarization is included in neutrino kinetic energy and mass.

\subsection{The common case of the neutrino passing through
the matter}

In \cite{14} a common method was developed  to avoid the above
paradox when from the very small energy arises huge mass change.
The example which we consider is very simple, therefore it will be
sufficient to take into account the law of momentum conservation
besides the law of energy conservation.
\par
Above we have considered the case when full energy of the matter
polarization caused by neutrino is included in the mass. If the
particle (neutrino) interaction with the matter is the left-right
symmetric one then the mass can be generated there (as it takes
place in strong and electromagnetic interactions). In this case we
have to share the full energy of the matter polarization caused by
the particle (neutrino) between the kinetic and mass parts of the
particle (neutrino) energy. We will suppose that the weak
interactions are the left-right symmetric ones and then we will
not consider the problem of mass generation in the weak
interactions.
\par
To solve this problem, it is necessary to compute full energy $W$
of the matter polarization and then taking into account the law of
energy-momentum conservation in the vacuum ($p, M$) and in the
matter ($p', M'$), - to distribute this full energy of
polarization between the kinetic and mass parts of the particle
(neutrino) energy. It coincides with the problem of polaron for a
certain interaction (for references see Wikipedia). So in the
matter
$$
E' = E + W , \eqno(20)
$$
and ($p_W = W v_{\nu}$)
$$
p' = p + p_W ,  \eqno(21)
$$
since $p^2 \gg M^2$ then neutrino is ultrarelativistic particle
and $v_{\nu} \simeq c$ (c is the light velocity) then $p_W \simeq
W$.
$$
p' \simeq p + W  . \eqno(22)
$$
Then the expression for neutrino energy in the matter will have
the following form:
$$
\sqrt{p^2 + M^2} + W = \sqrt{p'^2 + M'^2} \to
$$
$$
\to M'^2 + p'^2 = M^2 + 2 W \sqrt{p^2 + M^2} + W^2, \eqno(23)
$$
where $ p' \simeq p + W$. Then using expressions (20), (22) and
taking into account that $p^2 >> M^2$ from the expression. (23) we
obtain
$$
M'^2 -M^2 \simeq W p (\frac{M^2}{p^2}). \eqno(24)
$$
If to take into account that $p^2 >> M^2$ and $W \approx
10^{-12}$, then
$$
M'^2 \simeq M^2 + W p (\frac{M^2}{p^2}) \simeq M^2. \eqno(25)
$$
In this case the Wolfenstein's equation has the same form as
equation (1) since term $W$ originated from the left-right
symmetric interaction is inserted in this equation with the
left-right symmetric wave function.
$$
i \frac{d\nu_{Ph}}{dt} = (\sqrt{p'^2 + M'^2} \nu_{Ph} \to (p' \hat
I + \frac{\hat M'^2}{2p'}) \nu_{Ph} \to
$$
$$
\to (p' \hat I + \frac{(\hat M^2 + W p (\frac{M^2}{p^2})}{2p'})
\nu_{Ph} , \eqno(26)
$$
or taking into account that the term $W p (\frac{M^2}{p^2})$ is
very small
$$
i \frac{d\nu_{Ph}}{dt} =   ( p'I + \frac{{M}^2}{2p'}) \nu_{Ph},
\eqno(27)
$$
where $p' = (p + W)$. The expression for the neutrino transition
probability in this case has the following form:
$$
P(E', t, ...) = 1 - \sin ^{2} 2\theta'  \sin  {2\pi c t\over
L''_{o}} , \eqno(28)
$$
where $E' \simeq p'c$ and $ L''_{o} = {\sin 2 \theta'\over \sin
2\theta } L'_o$, since $M'^2 \simeq M^2$ then $ sin\theta \simeq
sin \theta'$, $ L''_{o} \simeq L'_o$
$$
L'_{0} = \frac{4 \pi E'_{\nu} \hbar}{\Delta m^2 c^3}, \quad
\sin^{2} 2\theta' = \sin^{2} 2\theta . \eqno(29)
$$
So, since changing of the neutrino effective mass is very small
then changing of the neutrino transition probability arises only
owing to the neutrino momentum change. It is necessary to take
into account that $p \gg W$, then this changing will be also very
small. We have come to the following conclusion: Taking into
account not only the law of neutrino  energy conservation but as
well as the law of neutrino momentum conservation, the neutrino
transition probability in the matter change is very small and
noticeable enhancement of the neutrino oscillations in the matter
does not appear (i.e., the condition (9) cannot be fulfilled). We
see that, the term which generates huge changing of the neutrino
effective mass and leads  to the resonance enhancement of neutrino
oscillations in the matter, appears since the law of momentum
conservation was not taken into account in the original
Wolfenstein's equation.

\section{Conclusion}

The Wolfenstein's equation is used to describe the neutrino
(particle) passing through the matter. Though this equation was
obtained to describe the neutrino passing through the matter (by
weak interactions which are left-side interactions) but it is a
Schrodinger's type of equation and therefore this equation is one
for left-right symmetric wave function and, correspondingly, it is
valid for the left-right symmetric interactions.
\par
It is supposed that while neutrino passing through the matter a
resonance enhancement of neutrino oscillations in the matter
appears. It is shown that Wolfenstein's equation, for neutrino
passing through the matter, contains a disadvantage (does not take
into account the law of momentum conservation). It leads, for
example, to changing of the effective mass of neutrino by the
value of $0.87 \cdot 10^{-2} eV$ from the very small value of the
energy polarization of the matter caused by neutrino which is
equal to $5 \cdot 10^{-12} eV$. After removing this disadvantage
(i.e., taking into account this law) we have obtained a solution
of this equation. In this solution a very small enhancement of
neutrino oscillations in the matter appears due to the smallness
of the energy polarization of the matter caused by neutrino.
\par
The author expresses his gratitude to Professors E. A. Kuraev, V.
B. Priesjev and I. V. Amirchanov for useful discussions.


\end{document}